\documentclass[a4paper,11pt]{article}
\pdfoutput=1 

\usepackage{jcappub} 

\usepackage{enumitem}

\title{GRB 170817A-GW170817-AT 2017gfo and the observations of NS-NS, NS-WD and WD-WD mergers}

\author[a,b,c]{J.~A.~Rueda,}
\author[a,b,c,d]{R.~Ruffini,}
\author[a,b]{Y.~Wang,}
\author[c,e]{U.~Barres~de~Almeida,}
\author[a,b]{C.~L.~Bianco,}
\author[a,b]{Y.C.~Chen,}
\author[a,b,f]{R.~V.~Lobato,}
\author[g]{C.~Maia,}
\author[a,b]{D.~Primorac,}
\author[a,b]{R.~Moradi}
\author[a,b]{and J.~F.~Rodriguez}

\affiliation[a]{Dipartimento di Fisica and ICRA, 
                     Sapienza Universit\`a di Roma, 
                     P.le Aldo Moro 5, 
                     I--00185 Rome, 
                     Italy}
                     
\affiliation[b]{ICRANet, 
                     P.zza della Repubblica 10, 
                     I--65122 Pescara, 
                     Italy}

\affiliation[c]{ICRANet-Rio, 
                     Centro Brasileiro de Pesquisas F\'isicas, 
                     Rua Dr. Xavier Sigaud 150, 
                     22290--180 Rio de Janeiro, 
                     Brazil}
                     
\affiliation[d]{Universit\'e de Nice Sophia Antipolis, 
                     CEDEX 2, Grand Ch\^{a}teau Parc Valrose, 
                     Nice, 
                     France}

\affiliation[e]{Centro Brasileiro de Pesquisas F\'isicas, 
                     Rua Dr. Xavier Sigaud 150, 
                     22290--180 Rio de Janeiro, 
                     Brazil}

\affiliation[f]{Departamento de F\'isica, Instituto Tecnol\'ogico de Aeron\'autica, ITA, S\~ao Jos\'e dos Campos, 12228-900 SP, Brazil}

\affiliation[g]{Instituto de F\'isica, Universidade de Bras\'ilia, 70910--900 Bras\'ilia, DF, Brazil}
\emailAdd{jorge.rueda@icra.it}
\emailAdd{ruffini@icra.it}

\date{\today}

\abstract{
The LIGO-Virgo Collaboration has announced the detection of GW170817 and has associated it with GRB 170817A. These signals have been followed after 11 hours by the optical and infrared emission of AT 2017gfo. The origin of this complex phenomenon has been attributed to a neutron star-neutron star (NS-NS) merger. In order to probe this association we confront our current understanding of the gravitational waves and associated electromagnetic radiation with four observed GRBs originating in binaries composed of different combinations NSs and white dwarfs (WDs). We consider 1) GRB 090510 the prototype of NS-NS merger leading to a black hole (BH); 2) GRB 130603B the prototype of a NS-NS merger leading to massive NS (MNS) with an associated kilonova; 3) GRB 060614 the prototype of a NS-WD merger leading to a MNS with an associated kilonova candidate; 4) GRB 170817A the prototype of a WD-WD merger leading to massive WD with an associated AT 2017gfo-like emission. None of these systems support the above mentioned association. The clear association between GRB 170817A and AT 2017gfo has led to introduce a new model based on on a new subfamily of GRBs originating from WD-WD mergers. We show how this novel model is in agreement with the exceptional observations in the optical, infrared, X- and gamma-rays of GRB 170817A-AT 2017gfo.
}

\begin{document}
\maketitle
\flushbottom

\section{Introduction}\label{sec:1}

The LIGO-Virgo Collaboration announced the detection of the gravitational-wave signal GW170817, at a luminosity distance of 40$^{+8}_{-14}$~Mpc, as consistent with the merging of a neutron star-neutron star (NS-NS) binary \cite{2017PhRvL.119p1101A}. The best-constrained parameter from the GW170817 data is the binary chirp mass, $\mathcal{M}\equiv (m_1 m_2)^{3/5}/M^{1/5} = 1.188^{+0.004}_{-0.002}~M_\odot$, where $m_1$ and $m_2$ are the binary merging components and $M = m_1 + m_2$ is the binary total mass. The 90\% confidence level of the total binary mass leads to the range $M = (2.73$--$3.29)~M_\odot$. The lowest value, i.e.~$M=2.73 M_\odot$, corresponds to the case of equal-mass components, $m_1=m_2 \equiv m = M/2 = 1.365 M_\odot$.

GW170817 has been associated with a short-duration gamma-ray burst (GRB) observed by the Gamma-ray Burst Monitor (GBM) on board the Fermi-satellite, GRB 170817A \cite{2017ApJ...848L..13A,2017ApJ...848L..14G}, as well as with the optical-infrared-ultraviolet ``kilonova'' emission, AT 2017gfo \cite{2017Sci...358.1556C}; see also \cite{2017Natur.551...64A,2017ApJ...848L..17C,2017ApJ...848L..18N}.

If the above associations were correct, then they would support the hypothesis that GW170817-GRB 1709817A-AT 2017gfo was produced by a NS-NS merger. The aim of this article is to gain additional insight into the nature of the electromagnetic signal of GRB 170817A by comparing and contrasting it with GRBs associated with four relevant systems: NS-NS mergers leading to a black hole (BH), NS-NS mergers leading to a more massive NS (MNS), NS-WD mergers leading to a MNS and WD-WD leading to a massive WD.

The article is organized as follows. In Sec.~\ref{sec:2} we recall the GRB subclasses associated with NS-NS and NS-WD mergers and we have introduced as a new subclass of GRBs originating from WD-WD mergers leading to a massive WD and their observational properties. In Sec.~\ref{sec:3} we analyze the gravitational-wave emission of NS-NS, NS-WD and WD-WD mergers. In Sec.~\ref{sec:4} we compare and contrast the X-ray and optical isotropic light-curves of GRB 090510A, GRB 130603B, GRB 060614 and GRB 170817A {including as well the infrared light-curve}. In Sec.~\ref{sec:5} we describe a new subclass of GRB originating from WD-WD merger. This model is an alternative mildly-relativistic uncollimated emission as oppose to the NS-NS ultra-relativistic merger. In Fig.~\ref{fig:wdwdejecta} we model the data of AT 2017gfo in the r, V, Ks and i bands following the WD-WD merger model. In Sec.~\ref{sec:6} we present the conclusions.

\section{GRB subclasses and observational properties of NS-NS and NS-WD mergers}\label{sec:2}

Short-duration GRBs are expected to be produced in the mergers of NS-NS and NS-BH binaries (see, e.g., refs.~\cite{1986ApJ...308L..47G,1986ApJ...308L..43P,1989Natur.340..126E,1991ApJ...379L..17N,1992ApJ...395L..83N}). Two different subclasses of short bursts from NS-NS mergers, depending on whether they lead to a MNS or to a BH, have been identified \cite{2015PhRvL.115w1102F,2015ApJ...808..190R,2016ApJ...832..136R}:
\begin{itemize}
\item
\emph{Authentic short GRBs (S-GRBs)}: they occur when the NS-NS merger leads to a BH \cite{2016ApJ...831..178R,2015ApJ...808..190R,2013ApJ...763..125M}. These bursts have $E_{\rm iso}\gtrsim 10^{52}$~erg and $E_{p,i}\gtrsim2$~MeV, and their electromagnetically inferred isotropic occurrence rate is $\rho_{\rm S-GRB}\approx \left(1.9^{+1.8}_{-1.1}\right)\times10^{-3}$~Gpc$^{-3}$~yr$^{-1}$ \cite{2016ApJ...832..136R, 2018ApJ...859...30R}. The distinct signature of the formation of a BH in a NS-NS merger in S-GRBs follows from the observations of the $0.1$--$100$~GeV emission by the \textit{Fermi}-LAT \cite{2018arXiv180207552R}, as observed in the best prototype case of GRB 090510. This is supported by the additional information of the following sources: GRB 090227B \cite{2013ApJ...763..125M}, GRB 140619B \cite{2015ApJ...808..190R}, GRB 090510 \cite{2016ApJ...831..178R} and more recently in GRB 081024B and GRB 140402A \cite{2017ApJ...844...83A}. The luminosity of the GeV emission follows a decreasing power-law with index $\gamma=-1.29\pm0.06$, when measured in the rest frame of the source \cite{2018arXiv180207552R}.

\item
\emph{Short gamma-ray flashes (S-GRFs)}: they occur when the NS-NS merger leads to a MNS; i.e. there is no BH formation \cite{2016ApJ...832..136R, 2018ApJ...859...30R}. These bursts have isotropic energy $E_{\rm iso}\lesssim10^{52}$~erg, peak energy $E_{p,i}\lesssim2$~MeV, and their electromagnetically inferred isotropic occurrence rate is $\rho_{\rm S-GRF}\approx 3.6^{+1.4}_{-1.0}$~Gpc$^{-3}$~yr$^{-1}$ \cite{2016ApJ...832..136R, 2018ApJ...859...30R}.


\end{itemize}

Besides the gamma-ray and X-ray emission, NS-NS mergers are expected to emit a kilonova in the infrared, optical and ultraviolet wavelengths, observable days after the merger \cite{1998ApJ...507L..59L,2010MNRAS.406.2650M,2013Natur.500..547T,2013ApJ...774L..23B}. This signal comes from the radioactive decay of $\sim 0.01~M_\odot$ r-process heavy material synthesized in the merger and it is expected to be nearly isotropic (see, e.g., refs.~\cite{1998ApJ...507L..59L,2010MNRAS.406.2650M}). The first kilonova associated with a short burst was established for GRB 130603B \cite{2013Natur.500..547T,2013ApJ...774L..23B}. With $E_{\rm iso}\approx 2\times 10^{51}$~erg \cite{2014ApJ...780..118F}, GRB 130603B belongs to the S-GRF subclass. The second association has been claimed for GRB 050709 \cite{2016NatCo...712898J} which, with an $E_{\rm iso}\approx 8\times 10^{49}$~erg, is also a S-GRF.

In addition to the above short bursts there is a subclass which show hybrid gamma-ray properties between long and short bursts. These gamma-ray flashes (GRFs) occur in a low-density circumburst medium (CBM), e.g.~$n_{\rm CBM}\sim∼10^{-3}$~cm$^{-3}$, and are not associated with supernovae (SNe) \cite{2006Natur.444.1050D,2016ApJ...832..136R, 2009A&A...498..501C}.

\begin{itemize}

\item
\emph{Gamma-ray flashes (GRFs)}: they are thought to originate in NS-WD mergers, see. e.g., \cite{2016ApJ...832..136R,2018ApJ...859...30R}. NS-WD binaries are notoriously common astrophysical systems \cite{2015ApJ...812...63C} and possible evolutionary scenarios leading to these mergers have been envisaged (see, e.g., refs.~\cite{1999ApJ...520..650F,2014MNRAS.437.1485L,2000ApJ...530L..93T}). These bursts, which show an extended and softer emission, have $10^{51}\lesssim E_{\rm iso}\lesssim 10^{52}$~erg, peak energy $0.2 \lesssim E_{p,i}\lesssim 2$~MeV, and their electromagnetically inferred isotropic occurrence rate is $\rho_{\rm GRF}=1.02^{+0.71}_{-0.46}$~Gpc$^{-3}$~yr$^{-1}$ \cite{2016ApJ...832..136R, 2018ApJ...859...30R}. This density rate appears to be low with respect to the number of estimated NS-WD binaries \cite{2015ApJ...812...63C} which can be of $(0.5$--$1)\times 10^4$~Gpc$^{-3}$~yr$^{-1}$ \cite{2009arXiv0912.0009T}. From the GRB side, we note that indeed only one NS-WD merger has been identified (see analysis of GRB 060614, in ref.~\cite{2009A&A...498..501C}). This implies that the majority of the expected mergers are under the threshold of the existing X- and gamma-ray detectors. A kilonova has been associated with GRB 060614 \cite{2015NatCo...6E7323Y}. Detail spectral and luminosity analysis has been presented in \cite{2009arXiv0912.0009T}. GRFs form a more massive NS and not a BH (see \cite{2016ApJ...832..136R} and references therein).

\end{itemize}

Finally, the case of GRB 170817A opens a new problematic: both its very low energetic $E_{\rm iso}\approx 5\times 10^{46}$~erg \cite{2017ApJ...848L..14G,2017ApJ...848L..12A,2017PhRvL.119p1101A} and the unprecedented details of the optical, infrared, radio and X-ray information of AT 2017gfo (see, e.g., refs.~\cite{2017Natur.551...64A,2017ApJ...848L..17C,2017ApJ...848L..18N}) which promise to give most precise information on the Lorentz factor of the different episodes underline the nature of this source. The observational features of GRB 170817A and AT 2017gfo lead us to consider here the possibility of an additional subclass of GRBs produced in WD-WD mergers leading to the formation of a massive WD.

\begin{itemize}
\item 
\emph{WD-WD mergers}:
The WD-WD merger rate has been recently estimated to be $(1$--$80)\times 10^{-13}$~yr$^{-1}$~$M_\odot^{-1}$ (at $2\sigma$) and $(5$--$9)\times 10^{-13}$~yr$^{-1}$~$M_\odot^{-1}$ (at $1\sigma$) \cite{2017MNRAS.467.1414M,2018MNRAS.476.2584M}. For a Milky Way-like stellar mass $6.4\times 10^{10}~M_\odot$ and using an extrapolating factor of Milky Way equivalent galaxies, 0.016~Mpc$^{-3}$ \citep{2001ApJ...556..340K}, it leads to a local cosmic rate $(0.74$--$5.94)\times 10^6$~Gpc$^{-3}$~yr$^{-1}$ ($2\sigma$) and $(3.7$--$6.7)\times 10^5$~Gpc$^{-3}$~yr$^{-1}$ ($1\sigma$). 

We are interested in WD-WD mergers forming as central remnant a massive ($\sim 1.2$--$1.5~M_\odot$), highly magnetized ($10^9$--$10^{10}$~G), fast rotating ($P=1$--$10$~s) WD (see \cite{2013ApJ...772L..24R} and references therein). Since there is no SN associated with GRB 170817A the merger should not lead to a SN explosion (see e.g. \cite{2018ApJ...857..134B} and references therein). This is therefore different, for instance, to the WD-WD mergers considered in \cite{2017NewA...56...84L} where the existence of a SN was explicitly envisaged. The above occurrence rate implies that $(12$--$22)\%$ of WD-WD mergers may explain the entire population of SN Ia. This is consistent with previous estimated rates of WD-WD mergers leading to SNe Ia (see e.g. \cite{2009ApJ...699.2026R}). The rest of the population is indeed sufficient to explain the population of GRB 170817A-like sources (see Sec.~\ref{sec:5}).

We show below in Sec.~\ref{sec:5} that the expected observables of these WD-WD mergers in the X- and gamma-rays, and in the optical and in the infrared wavelengths are suitable for the explanation of the GRB 170817A-AT 2017gfo association. 

We recall that the name kilonova was coined in \cite{2010MNRAS.406.2650M} to the optical transient produced in NS-NS mergers in view of their optical luminosity, $\sim 10^{41}$~erg~s$^{-1}$, which is approximately 1000 times the one of novae, $\sim 10^{38}$~erg~s$^{-1}$. The results of this work show that indeed this designation can be extended to the optical transient produced in WD-WD mergers. In this case the kilonova is not powered by the decay of r-process material but by the energy released by accretion onto the new WD formed in the merger (see Sec.~\ref{sec:5}).

In addition, it is interesting that the above mentioned physical properties (e.g. mass, rotation period and magnetic field) of the WD formed in the merger process correspond to the ones described by the WD model of soft gamma-repeaters (SGRs) and anomalous X-ray pulsars (AXPs) \cite{2012PASJ...64...56M,2013ApJ...772L..24R}. Indeed, the WD-WD merger rate is high enough to explain the Galactic population of SGRs/AXPs. Therefore the possible evolution of GRB 170817A-AT 2017gfo into this kind of sources has to be attentively scrutinized in the forthcoming months (see Sec.~\ref{sec:5}).

\end{itemize}

\section{Gravitational-wave emission of NS-NS, NS-WD and WD-WD mergers}\label{sec:3}

We first compare and contrast the gravitational-wave emission expected from the above GRBs originating from mergers and their observed electromagnetic emissions with the ones associated with GW170817, namely GRB 170817A and AT 2017gfo \cite{2017PhRvL.119p1101A}.

We here consider our canonical model of GRBs assuming uncollimated emission, in absence of observational evidence of an achromatic jet break (see \cite{2018ApJ...859...30R,2018ApJ...852...53R} and references therein). We understand that a vast literature exists on alternative models based on a large variety of structured beamed emission which we do not consider here in view of the above considerations. A different approach leading to an agreement with the observational data is here proposed.

\begin{figure}
\centering
\includegraphics[width=\hsize,clip]{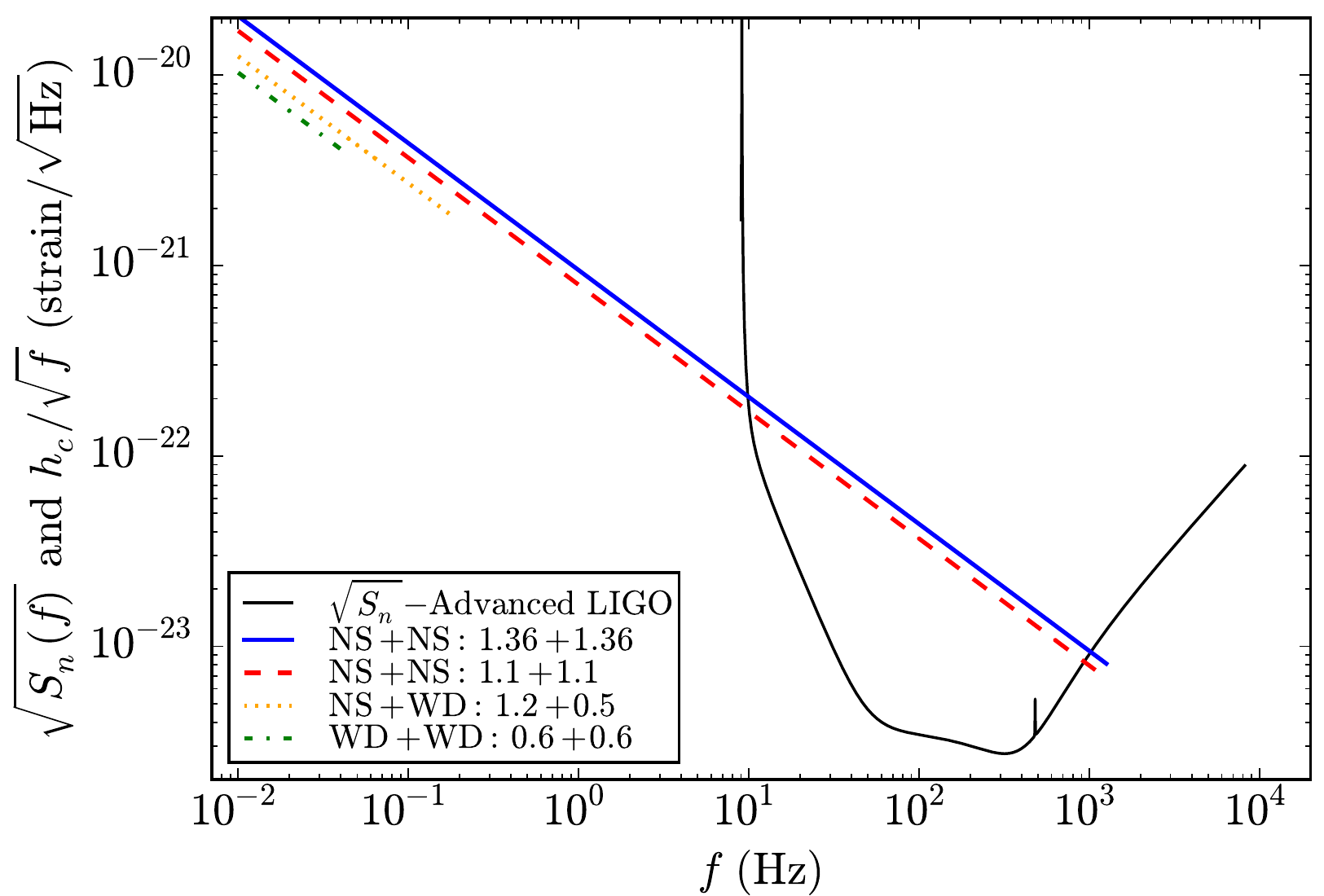}
\caption{Source ASD, $h_c(f)/\sqrt{f}$, together with the one-sided ASD of the Advanced LIGO detector's noise, $\sqrt{S_n}$, for representative examples of S-GRBs (GRB 090510A-like), S-GRFs (GRB 130603B-like) and GRFs (GRB 060614-like). We have also included the expected ASD for a representative WD-WD binary. For the sake of the comparison all sources have been artificially assumed to be at a luminosity distance of 40~Mpc (cosmological redshift $z\approx 0.009$). For details of the gravitational-wave emission of these binaries see \cite{2018ApJ...859...30R}.}\label{fig:hc}
\end{figure}

In Fig.~\ref{fig:hc} we show the gravitational-wave source amplitude spectral density (ASD), $h_c(f)/\sqrt{f}$, together with the one-sided ASD of the Advanced LIGO detector's noise, $\sqrt{S_n}$ \cite{2016LRR....19....1A}. The gravitational-wave characteristic strain is $h_c = (1+z)\sqrt{(1/10) (G/c^3) dE/df_s}/d_L(z)$, where $d_L(z)$ is the luminosity distance to the source, $f=f_s/(1+z)$ and $f_s$ are the gravitational-wave frequency in the detector's and in the source's frame and $dE/df_s$ is the gravitational-wave spectrum, respectively. For the luminosity distance we adopt a $\Lambda$CDM cosmology with $H_0 = 71$~km~s$^{-1}$, $\Omega_M = 0.73$ and $\Omega_\Lambda=0.23$ \cite{2015ApJ...802...20R}. The spectrum of the binary inspiral can be adopted from the traditional quadrupole formula, $dE/df_s = (2^{1/3}/3) (\pi G)^{2/3} \mathcal{M}^{5/3} f_s^{-1/3}$, where we recall $\mathcal{M}\equiv (m_1 m_2)^{3/5}/M^{1/5}$ and $M = m_1+m_2$ are, respectively, the chirp mass and total mass of the binary. We cut the gravitational-wave emission of the inspiral at the point where the two stars enter into contact, namely at a distance $r = R_1 + R_2$, where $R_1$ and $R_2$ are the stellar radii. For the NS radii we use the mass-radius relation shown in \cite{2015PhRvD..92b3007C} obtained with the GM1 equation of state while, for the WDs, we use the mass-radius relation in \cite{2011PhRvD..84h4007R} obtained with the relativistic Feynman-Metropolis-Teller equation of state.

To represent the emission of a S-GRB we adopt the parameters of GRB 090510A, the first identified NS-NS merger leading to a BH \cite{2016ApJ...831..178R}. We thus use $m_1 = m_2 = 1.36~M_\odot$, consistent with the condition that the merging mass exceeds the NS critical mass in the case of the GM1 nuclear equation of state which, for a non-rotating NS, is $M_{\rm crit}\approx 2.4~M_\odot$ \cite{2015PhRvD..92b3007C}. We are here neglecting, for simplicity, the angular momentum distribution of the post-merger configuration which would lead to a more complex analysis of the initial merging masses leading to BH formation (Rodriguez, et al., in preparation).

For a S-GRF, we assume a GRB 130603B-like source with $m_1 = m_2 = 1.1~M_\odot$, consistent with the condition that the merged object is a massive but stable NS{, consistent with the adopted NS critical mass value}. 

For a GRF, we adopt a GRB 060614-like source, namely a NS-WD binary with $m_1 = 1.2~M_\odot$ and $m_2 = 0.5~M_\odot$ \cite{2009A&A...498..501C}.

To compare and contrast the gravitational-wave emission, we have located all the sources at a distance of $d = 40$~Mpc, as the one of GW170817 \cite{2017PhRvL.119p1101A}. The gravitational-wave emission associated with the inspiral phase of the NS-NS mergers (GRB 090510A-like and GRB 130603B-like) would be consistent, both in the characteristic strain and the spanned frequency range, with the ones of GW170817 (see ref.~\cite{2017PhRvL.119p1101A}, for details on this source). Instead, the emission of NS-WD and WD-WD mergers is outside the Advanced LIGO frequency band. For details of the gravitational-wave emission from these binaries see \cite{2018ApJ...859...30R}.

\section{Comparison of the prompt, X-rays and optical light-curves of NS-NS and NS-WD mergers}\label{sec:4}

We turn now to the comparison of the electromagnetic emission of all these binaries, namely of S-GRBs, S-GRFs and GRFs, with the ones of GRB 170817A - AT 2017gfo.

--{\bf\emph{GeV emission}}.

\begin{figure}
    \centering
    \includegraphics[width=\hsize,clip]{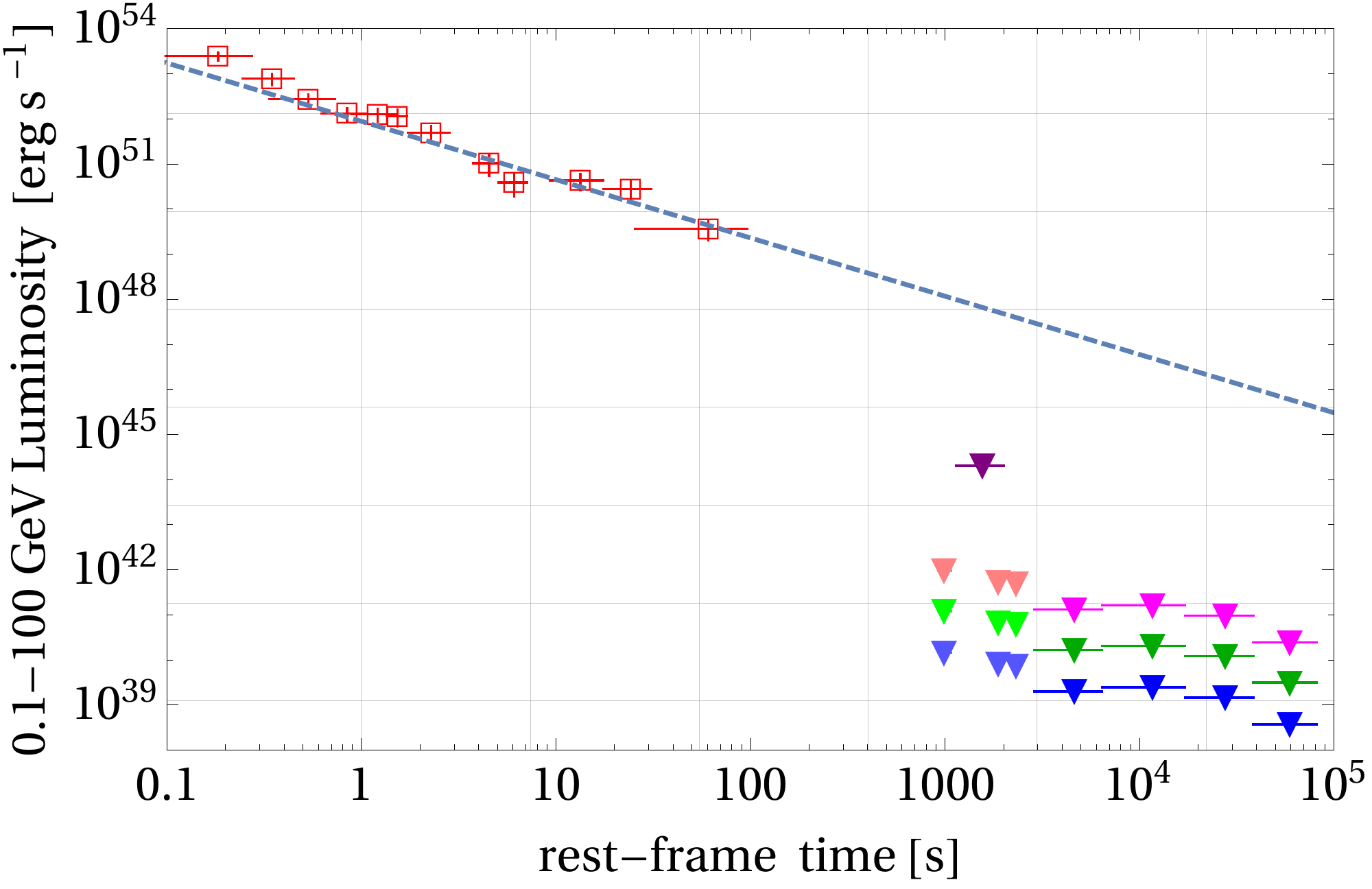}
    \caption{Comparison of the rest-frame, $0.1$--$100$~GeV band light-curve of GRB 090510A (red empty squares) with the corresponding observational upper limits of GRB 170817A (filled triangles). The dashed line indicates the power-law decay with slope index $-1.29\pm 0.06$, observed in S-GRBs (see \cite{2018arXiv180207552R} for details, and references therein). The purple triangle indicates the upper limit by Fermi-LAT of $9.7\times 10^{43}$~erg~s$^{-1}$ from $1151$ to $2025$~s from the Fermi trigger time. AGILE upper limits (pink, green and blue triangles) were calculated through extrapolation to the Fermi-LAT working energy band ($0.1$--$100$~GeV). We assume the spectral indices $-2.0$ (blue triangles), $-1.0$ (green triangles), $-0.1$ (pink triangles). The data of GRB 170817A were retrieved from \cite{2017arXiv171005450F, 2017ApJ...850L..27V}.}
    \label{fig:GeV}
\end{figure}

\begin{figure}
\centering
\includegraphics[width=\hsize,clip]{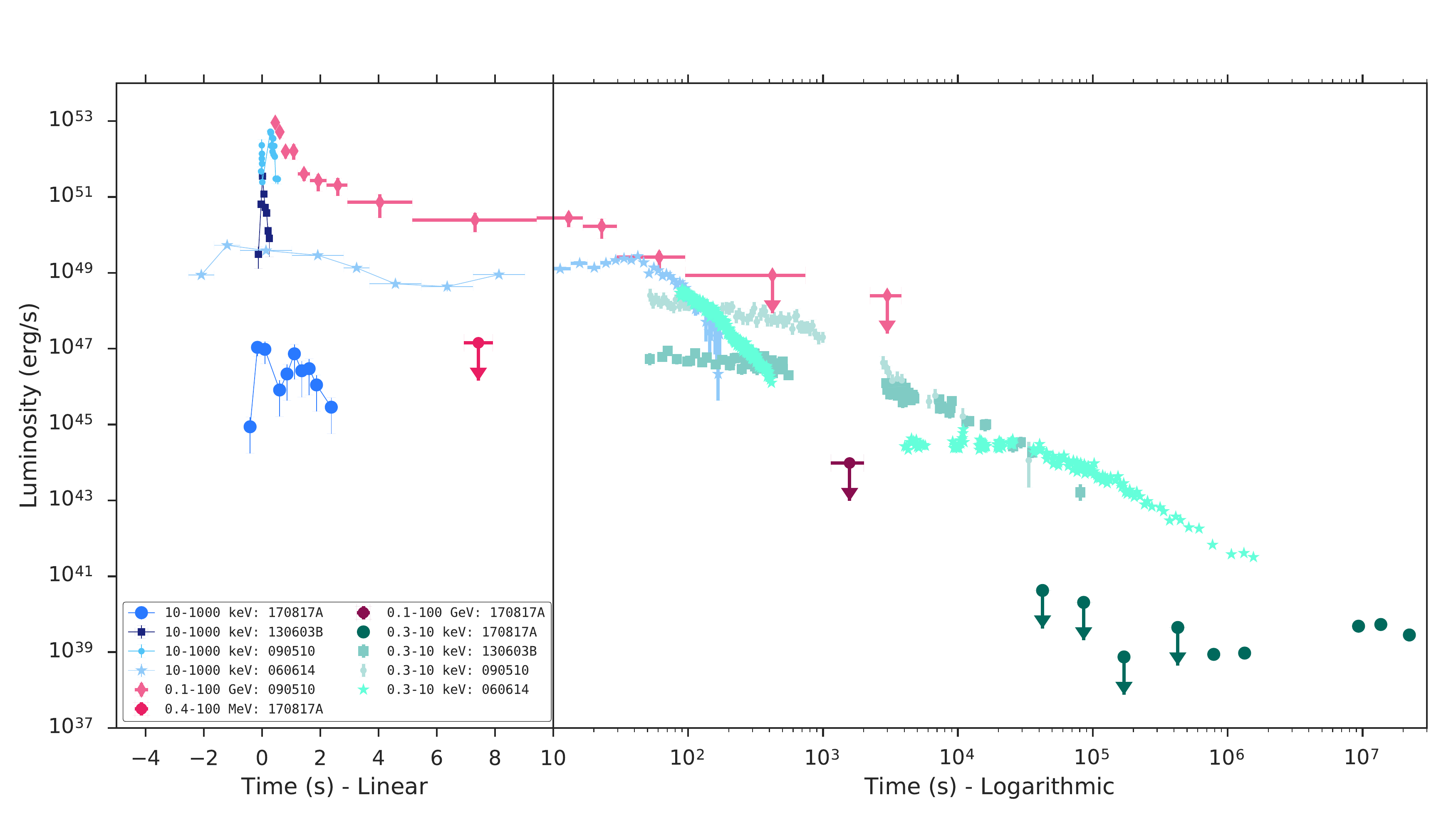}
\caption{Light-curves of GRBs 060614, 090510A, 130603 and 170817A in the cosmological rest-frame. We show the gamma-ray ($10$--$1000$~keV) prompt and the X-ray  ($0.3$--$10$ keV) emissions. The first 10 seconds are plotted in a linear scale and longer times in the logarithmic scale.}\label{fig:promptandX}
\end{figure}

\begin{figure}
\centering
\includegraphics[width=\hsize,clip]{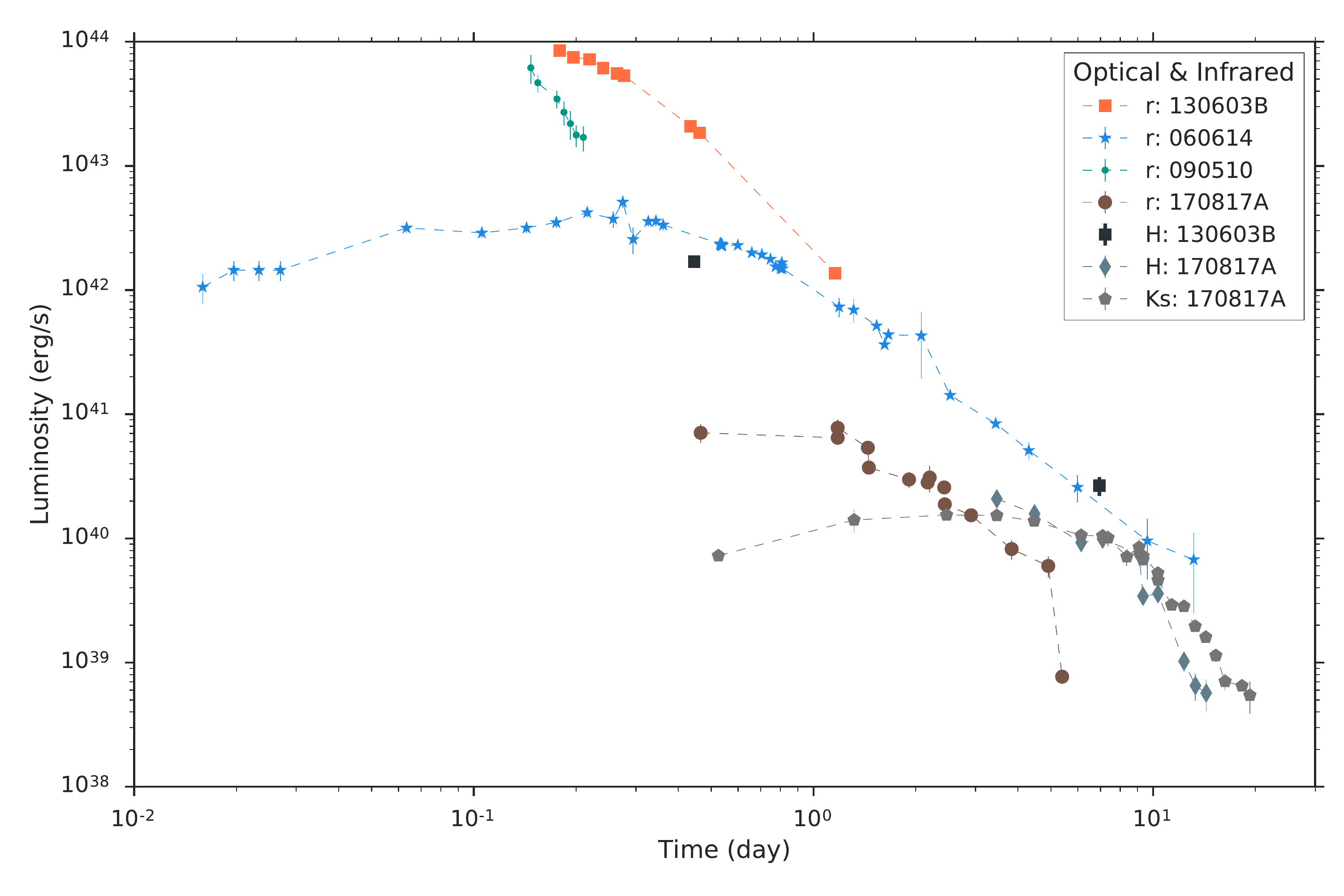}
\caption{Optical (r band) and infrared (H and Ks bands) light-curves of GRBs 060614, 090510A, 130603 and 170817A in the cosmological rest-frame.}\label{fig:optical}
\end{figure}

A first general conclusion can be directly inferred for the absence in GRB 170817A of the GeV emission (see e.g.~\cite{2017arXiv171005450F, 2017ApJ...850L..27V}): we can at once conclude that GRB 170817A is not consistent with a S-GRB, a NS-NS merger leading to a BH formation (see Fig.~\ref{fig:GeV} and  \cite{2018arXiv180207552R}). This conclusion is in agreement with the one obtained from the analysis of the gamma-ray prompt emission and the X-ray emission; see below and Figs.~\ref{fig:promptandX}--\ref{fig:optical} and Tables~\ref{tab:prompt}--\ref{tab:binaries}.

--{\bf\emph{Gamma-ray prompt emission}}.

Figure~\ref{fig:promptandX} shows the gamma-ray ($10$--$1000$~keV) prompt emission isotropic rest-frame light-curves of GRBs 090510, 130603B, 060614 and 170817A. In Table~\ref{tab:prompt} we compare and contrast the isotropic energy, peak luminosity and peak energy of the prompt emission for the same GRBs.

We can conclude that the gamma-ray prompt emission from GRB 170817A is not consistent with the one observed in GRBs 090510A, 130603B and 060614.

\begin{table*}
\centering
\small\addtolength{\tabcolsep}{-1.0pt}
\begin{tabular}{lcccccc}
\hline\hline
& GRB& $E_{\rm iso}$~(erg) &$L_{\rm peak}$ (erg~s$^{-1}$) & $E_{\rm peak}$ (MeV) & Reference\\
\hline
 &060614& $2.17\times 10^{51}$ & $ 3\times 10^{49}$ & $0.34^{+0.24}_{-0.1} $ & GCN Circular 5264 \cite{2006GCN..5264....1G} \\
& 090510 & $3.95\times 10^{52}$ & $9.1\times 10^{52}$ & $7.89 \pm 0.76$ & \cite{2016ApJ...832..136R}\\
& 130603B & $ 2.1\times 10^{51}$ & $4.1\times 10^{51}$ & $0.90 \pm 0.14$ & GCN Circular 14771 \cite{2013GCN.14771....1G} \\ 
& 170817A & $ 3.1\times 10^{46}$ & $1.2\times 10^{47}$ & $0.082 \pm 0.021$ & GCN Circular 21520 \cite{2017GCN.21520....1V} \\
\hline 
\end{tabular}
\caption{The isotropic energy, luminosity of the peak and peak energy of GRBs 060614, 090510, 130603B and 170817A in the prompt gamma-ray emission phase.}
\label{tab:prompt}
\end{table*}

--{\bf\emph{X-rays}}.

In addition, we also show in Fig.~\ref{fig:promptandX} the corresponding X-ray isotropic light-curves, in the rest-frame $0.3$--$10$~keV energy band. It can be seen the overlapping of the light-curves at times $t\gtrsim 5000$~s from the BAT trigger \cite{2013GCN.14735....1M}. We recall that we had presented a first comparison of the X-ray light-curves of GRBs 090510A and 130603B in \cite{2013GCN.14913....1R}. The match of the X-ray light-curves occurs irrespectively of their isotropic energies which differ up to a factor of $\approx 20$ for instance in the case of GRB 130603B, $E_{\rm iso} = 2.1\times 10^{51}$~erg \cite{2013GCN.14772....1F} and GRB 090510A, $E_{\rm iso} = 3.95\times 10^{52}$~erg \cite{2016ApJ...831..178R}).

We can see that the X-ray emission from GRB 170817A is not consistent with the one observed in GRBs 090510A, 130603B and 060614.

--{\bf\emph{Optical and infrared}}.

We show in Fig.~\ref{fig:optical} the optical (r band) and and infrared (H and Ks bands)light-curve of GRB 090510A \cite{2010ApJ...709L.146D,2012A&A...538L...7N}, GRB 130603B \cite{2013Natur.500..547T,2013ApJ...774L..23B}, GRB 060614 \cite{2009A&A...498..501C,2015NatCo...6E7323Y}, and GRB 170817A (i.e. AT 2017gfo~\cite{2017Sci...358.1556C,2017Natur.551...64A,2017ApJ...848L..17C,2017ApJ...848L..18N}).

Since a kilonova emission had been associated with GRB 130603B \cite{2013Natur.500..547T}, the similarity of this GRB in the X-rays with GRB 090510A boosted us to seek for a kilonova signature in the case GRB 090510A. This appears to be confirmed by Fig.~\ref{fig:optical}. We can conclude that all of these sources, S-GRFs, S-GRBs and GRFs, can produce a kilonova emission in line with the source AT 2017gfo, the kilonova associated with GRB 170817A.

Some of the above correlations of the electromagnetic emissions clearly originate from the traditional kilonova models based on ultra-relativistic regimes in NS-NS and NS-WD mergers pioneered in \cite{Lattimer:1977igd,1998ApJ...507L..59L,1989Natur.340..126E}, following the classical work of \cite{1980ApJ...237..541A,1982ApJ...253..785A}.

The common asymptotic behavior at late times of the X-ray and optical emission of GRB 090510A \cite{2013GCN.14913....1R} and GRB 130603B \cite{2013Natur.500..547T} are a manifestation of a common synchrotron emission as recently outlined in \cite{2017arXiv171205000R}. This approach is also supported by the parallel behavior of the optical and X-ray emission of GRB 060614 (see Figs.~\ref{fig:promptandX} and \ref{fig:optical}). Similar behavior was already indicated in the case of long GRBs (BdHNe) \cite{2014A&A...565L..10R,2015ApJ...798...10R}. In fact, it is interesting to use this similarity to set a lower limit to the mass of the expanding matter by equating the energy emitted in the X-rays, $E_{\rm iso,X}$ (see Table~\ref{tab:binaries}), to the kinetic energy, $E_{\rm kin} = (\Gamma-1) Mc^2.$, where $\Gamma$ is the Lorentz factor. By using for instance $\Gamma \sim 1.2$ ($v/c\sim 0.5$), a typical value obtained for the expanding blackbody component in the X-ray afterglow of BdHNe \cite{2018ApJ...852...53R}, we obtain $9.2\times 10^{-4}~M_\odot$, $9.4\times 10^{-5}~M_\odot$ and $2.4\times 10^{-4}~M_\odot$ for GRBs 090510A, 130603B and 060614. These lower limits are indeed in line with the ejecta mass obtained from an independent analysis of the optical emission (see e.g. \cite{2013Natur.500..547T,2013ApJ...774L..23B,2016NatCo...712898J,2010MNRAS.406.2650M}). The same analysis of the X-ray emission applied to GRB 170817A would lead to a nearly ten orders of magnitude less massive ejecta, giving further support to a possible different nature of this source.

The above contrast of the gamma- and X-rays observational properties of GRB 170817A with respect to GRBs produced by NS-NS and NS-WD mergers has led us to consider the possibility of an additional subclass of GRBs produced by the merger of a still different compact-star binary, a WD-WD merger, leading to a massive WD. We proceed now to discuss the framework of such a model to explain the electromagnetic properties of the association GRB 170817A - AT 2017gfo.

\section{{WD-WD mergers as an alternative mildly relativistic uncollimated emission for GRB 170817A-AT 2017gfo}}\label{sec:5}

The estimated WD-WD merger rate (see Sec.~\ref{sec:2}) implies that $0.1\%$ of WD-WD mergers can explain the GRB 170817A-like population for which a lower limit of $(30$--$630)$~Gpc$^{-3}$~yr$^{-1}$ has been recently obtained (see \cite{2018NatCo...9..447Z} for details).

The energy observed in gamma-rays in GRB 170817A, $E_{\rm iso}\approx 3 \times 10^{46}$~erg, can originate from flares owing to the twist and stress of the magnetic field lines during the merger process: a magnetic energy of $2\times 10^{46}$~erg is stored in a region of radius $10^9$~cm and magnetic field of $10^{10}$~G \cite{2012PASJ...64...56M}.

\begin{figure}
\centering
\includegraphics[width=\hsize,clip]{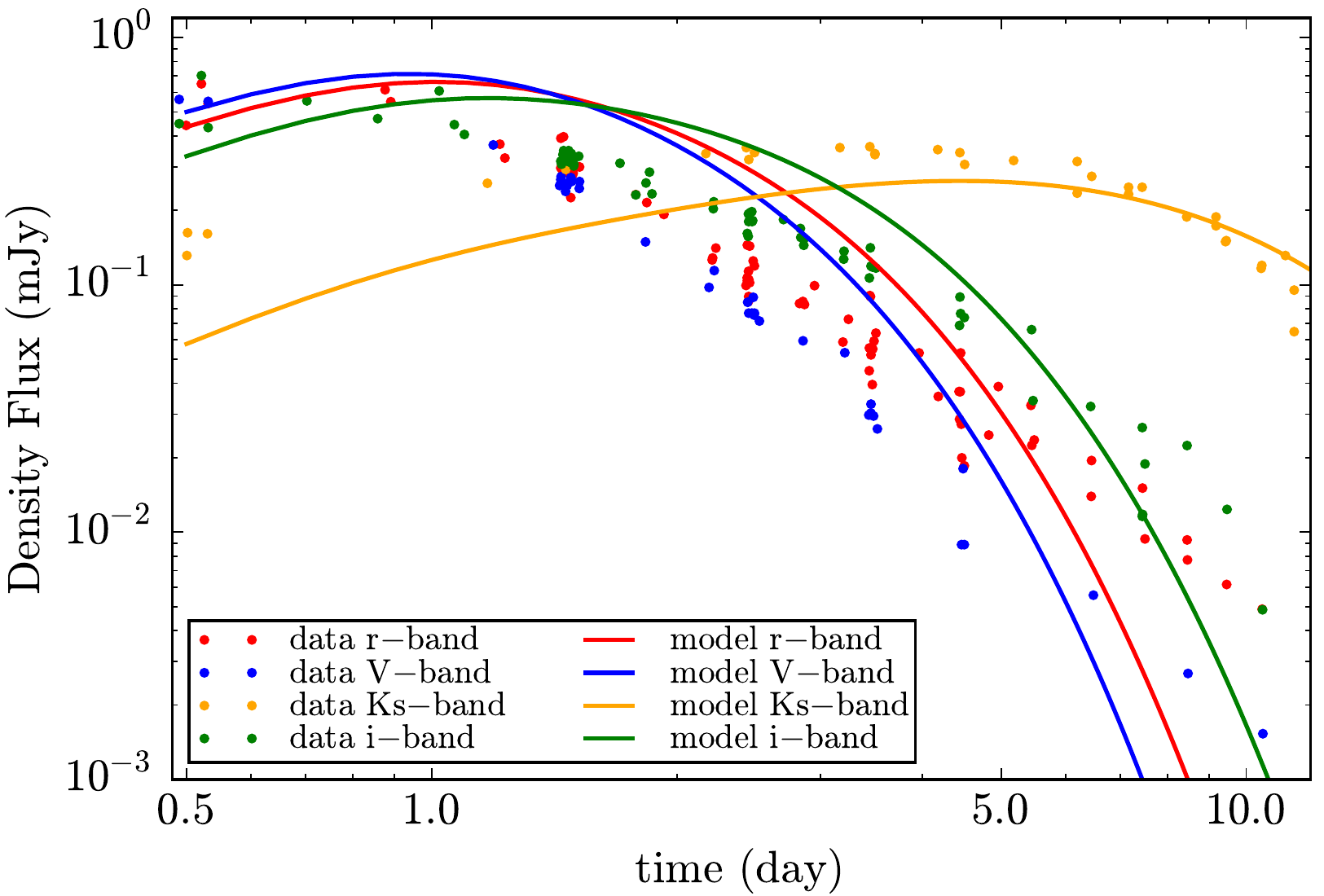}
\caption{Points: observed optical and infrared density flux of AT 2017gfo~\cite{2017ApJ...848L..17C,2017ApJ...848L..18N}. Solid curves: corresponding theoretical expectation from the cooling of $10^{-3}\,M_\odot$ of WD-WD merger ejecta heated by fallback accretion onto the newly-formed central WD.}\label{fig:wdwdejecta}
\end{figure}

The emission at optical and infrared wavelengths (see Fig.~\ref{fig:wdwdejecta}), can be explained from the adiabatic cooling of $10^{-3}\,M_\odot$ ejecta from the merger \cite{2009A&A...500.1193L,2011ApJ...737...89D} heated by fallback accretion onto the newly-formed WD \cite{2009A&A...500.1193L}. The ejecta becomes transparent at times $t~\sim 1$~day with a peak bolometric luminosity of $L_{\rm bol}\sim 10^{42}$~erg~s$^{-1}$. The fallback accretion injects to the ejecta $10^{47}$--$10^{49}$~erg~s$^{-1}$ at early times and fall-off following a power-law behavior (see \cite{2009A&A...500.1193L} for details). The kilonovae from WD-WD mergers are therefore powered by a different mechanism with respect to the kilonovae from NS-NS mergers which are powered by the radioactive decay of r-process heavy material.

At times $t\sim 100$--$200$~day, the ejecta are expected to become transparent to the X-rays leading to a luminosity of $\approx 10^{39}$~erg~s$^{-1}$ as the one recently observed in GRB 170817A (see Fig.~\ref{fig:promptandX}). At earlier times, the X-rays from fallback accretion are instead absorbed by the ejecta and are mainly transformed into kinetic energy then increasing the expansion velocity of the ejecta, from an initial non-relativistic value $0.01~c$ typical of the escape velocity from the WD, to a mildly relativistic velocity $0.1~c$.This mildly relativistic velocity is also consistent with the value derived from the evolution of blackbody spectra observed from $\sim 0.5$ day to $\sim 7$ day. We present the detailed analysis of all the above properties of WD-WD mergers in \cite{2018arXiv180707905R}.

As we have mentioned in Sec.~\ref{sec:2}, the WD formed in the merger can become an SGR/AXP \cite{2012PASJ...64...56M,2013ApJ...772L..24R}. Thus, there is the possibility that, if a WD-WD merger produced GRB 170817A-AT 2017gfo, an SGR/AXP (a WD-pulsar) will show in this sky position in the near future.

The observational features of these WD-WD mergers in the X- and the gamma-rays as well as in the optical and infrared wavelengths are an important topic by their own. We are going to present elsewhere additional details on the multiwavelength observables of the merger, of the post-merger and of the ejected matter.

\section{Discussion and conclusions}\label{sec:6}

\begin{table*}
\centering
\small\addtolength{\tabcolsep}{-1.0pt}
{\tiny
\begin{tabular}{lccccccccc}
\hline\hline
Subclass  &  \emph{In-state} $\to$ \emph{Out-state} &  $T_{90}$  &  $E_{\rm p,i}$ &  $E_{\rm iso}$ & $E_{\rm iso,Gev}$ & $E_{\rm iso,X}$ & Kilonova & $L_{\rm p,KN}$ & Hz-kHz GW\\
& &  (s) & (MeV) & (erg) & (erg) & (erg) & & (erg~s$^{-1}$) & \\
\hline
S-GRFs & NS-NS $\to$ MNS  &  $\lesssim2$  &  $\lesssim2$ &  $10^{49}$--$10^{52}$ &  $-$ & $10^{49}$--$10^{51}$ & RPKN & $10^{41}$ & Yes \\
S-GRBs & NS-NS $\to$ BH  &  $\lesssim2$  &  $\gtrsim2$ &  $10^{52}$--$10^{53}$ & $10^{52}$--$10^{53}$ & $\lesssim 10^{51}$ & RPKN & $10^{41}$ & Yes \\
GRFs  & NS-WD $\to$ MNS &  $2$--$100$  &  $0.2$--$2$ & $10^{51}$--$10^{52}$ & $-$ & $10^{49}$--$10^{50}$ & TBD & $10^{41}$ & No\\
GR-K  & WD-WD $\to$ MWD &  $\lesssim 1$  &  $\lesssim 0.1$ &  $\lesssim 10^{47}$ & $-$ &  $10^{38}$--$10^{39}$ & FBKN & $10^{41}$ & No\\
\hline
\end{tabular}
}
\caption{Summary of the GRB subclasses discussed in this article: short gamma-ray flashes (S-GRFs), authentic short GRBs (S-GRBs) and gamma-ray flash kilonovae (GR-K) introduced in this work. See further details in \cite{2016ApJ...832..136R,2018ApJ...859...30R,2018arXiv180707905R} and references therein. The columns indicate the GRB subclass, \emph{in-states} (progenitor) and \emph{out-states} (final outcome), the $T_{90}$ in the rest-frame, the rest-frame spectral peak energy $E_{\rm p,i}$ and $E_{\rm iso}$ (rest-frame $1$--$10^4$~keV), the isotropic energy of the GeV emission $E_{\rm iso,GeV}$ (rest-frame $0.1$--$100$~GeV), the isotropic energy of the X-ray data $E_{\rm iso,X}$ (rest-frame $0.3$--$10$~keV), the presence or not of kilonova and its nature (radioactive-powered kilonova, RPKN; fallback-powered kilonova, FBKN), the typical associated kilonova peak luminosity, and finally the presence or not of Hz-kHz gravitational-wave emission, e.g. as the one detectable by LIGO-Virgo. As for the kilonova from NS-WD mergers we have specified ``to be defined (TBD)'' since this possibility has not been yet explored in the literature.}\label{tab:binaries}
\end{table*}

In this work we have compared and contrasted the gravitational-wave and the electromagnetic emission of canonical GRBs associated with NS-NS (both S-GRBs and S-GRFs), NS-WD (GRFs) and WD-WD mergers with the one of the associated sources GRB 170817A-GW170817-AT 2017gfo. We present in Table~\ref{tab:binaries} a summary. As a canonical S-GRB we use GRB 090510A, for S-GRF we use GRB 130603B and for GRF we use GRB 060614 and for GRB 170817 a twin WD with component masses $M= 0.6 M_\odot$, see Fig.~\ref{fig:hc}. 

We can conclude:

\begin{enumerate}

\item 
The comparison of the properties of NS-NS (S-GRFs and S-GRBs) and NS-WD (S-GRFs) with GRB 170817A-GW170817-AT 2017gfo shows that all of them may include a kilonova (AT 2017gfo-like). Only in NS-NS there could be the Hz-kHz gravitational-wave emission needed to explain the energetics of GW170817 (see Fig.~\ref{fig:hc}). However, this solution necessarily implies an X and gamma-ray emission in the prompt phase that is missing in the case of GRB 170817A (see data up to 10~s in Fig.~ \ref{fig:promptandX}). Indeed, the observational features in gamma- and X-rays of GRB 170817A contrasts with any other GRB associated with the above binary progenitors (see Figs.~\ref{fig:promptandX}--\ref{fig:optical} and Tables~\ref{tab:prompt}--\ref{tab:binaries}). In conclusion, the NS-NS scenario cannot explain the association GRB170817A-GW170817.

\item 
Indeed, the X-ray and gamma-ray observations of GRB 170817A clearly show that we are in presence of a phenomenon much less energetic than the one observed in any S-GRB, S-GRF, GRF, or BdHN, and perhaps with a substantially larger occurrence rate (see e.g.~\cite{2018NatCo...9..447Z}). These observations have led us to consider a new subclass of GRBs, also with binary progenitors, originating from WD-WD mergers leading to a massive WD. The occurrence rate of these mergers can explain the rate of GRB 170817A-like sources, they produce a gamma- and X-ray emission consistent with the ones observed in GRB 170817A and cannot be associated with GW170817. 

\item
The optical and infrared emission AT 2017gfo can be powered by a different physical mechanism with respect to the radioactive decay of r-process heavy material synthesized in the much more energetic NS-NS mergers: it can be alternatively explained by the cooling of the ejecta expelled in a WD-WD merger and heated up by fallback accretion onto the newly-formed massive WD, see Fig.~\ref{fig:wdwdejecta}. In view of the above difference we propose to call radioactive-powered kilonovae (RPKNe) the optical transient produced by NS-NS mergers and fallback-powered kilonovae (FBKNe) the one by WD-WD mergers. See Table~\ref{tab:binaries}.

\item
The ejecta from a WD-WD merger are different from the ejecta from a NS-NS merger in that they have a lighter nuclear composition with respect to the one of the ejecta of NS-NS mergers which is made of r-processed heavy nuclei. The spectroscopic identification of atomic species can therefore discriminate between the two scenarios. However, such an identification has not been possible in any of the observed kilonovae since it needs accurate models of atomic spectra, nuclear reaction network, density profile, as well as radiative transport (opacity) that are not yet available.

\item
These WD-WD mergers opens the possibility to a new subclass of GRBs with a much less energetic and softer prompt emission whose observation would benefit from a new mission operating in soft X-rays like, e.g., THESEUS \cite{2017arXiv171004638A}. In addition, as we have shown, the outcome of such a GRB, namely a massive, highly magnetized, fast rotating WD, may become in due time observable as SGRs/AXP.

\item
Since the early submission of our paper additional observations in the optical and in the X-rays have appeared (see e.g.~\cite{2018A&A...613L...1D,2018NatAs...2..751L}) which have allowed to strength our conclusions (see e.g. Fig.~\ref{fig:wdwdejecta} and \cite{2018arXiv180707905R}). What we can do at this stage from a theoretical point of view is to formulate what conventional physics can tell us about these events and this has been done in this article.

\item
It is clear that the only possibility of a null chance coincidence of GRB 170817A and GW170817 is to assume that one of the events does not exist in reality. If the two events exist then there is non-null chance coincidence by definition and its evaluation has been estimated (see e.g.~\cite{2017ApJ...848L..13A}). The association of these events, from an observational point of view is, in our opinion, not yet sufficiently established to formulate a well-motivated answer. It is auspicable, as soon as the LIGO collaboration releases the templates of the gravitational-wave source GW170817 in the interferometers, to reconstruct the precise chronology of the space-time sequence of events in the LIGO detectors and in the Fermi and Integral satellites, necessary to validate the association between GW170817 and GRB 170817A.

\end{enumerate}

\acknowledgments
We thank P.~Lor\'en-Aguilar for discussions on WD-WD mergers. We thank both the Referee and Scientific Editor for valuable help in optimizing the presentation of our paper as well as for the detailed and constructive discussions.

\bibliographystyle{JHEP}
\bibliography{references.bib}

\end{document}